\documentclass[a4paper,scriptaddress,onecolumn,prl,12pt]{revtex4}
\usepackage{amsmath}
\usepackage{makeidx}
\usepackage{amsfonts}
\usepackage{graphicx}
\usepackage[ansinew]{inputenc}
\usepackage[usenames,dvipsnames]{pstricks}
\usepackage{subfigure}
\usepackage{epsfig}
\usepackage{pst-grad}
\usepackage{pst-plot}
\usepackage[colorlinks,hyperindex]{hyperref}
\usepackage[dvips]{geometry}

\hypersetup{
colorlinks,
citecolor=blue,
linkcolor=black,
urlcolor=black,
}

\DeclareMathOperator{\sech}{sech}
\makeindex

\begin{document}

\title{On the study of oscillons in scalar field theories: A new approach}
\author{R. A. C. Correa$^{1,2,}$\footnote{
rafael.couceiro@ufabc.edu.br} and A. de Souza Dutra$^{2,}$\footnote{
dutra@feg.unesp.br}\medskip }
\affiliation{$^{1}$Centro de Ciências Naturais e Humanas, Universidade Federal do ABC,
09210-580, Santo André, SP, Brazil\medskip \\
$^{2}$São Paulo State University-UNESP-Campus de Guaratinguetá-DFQ, Av. Dr.
Ariberto Pereira Cunha, 333, 12516-410, Guaratinguetá, SP, Brazil}

\begin{abstract}
In this work we study configurations in one-dimensional scalar field theory,
which are time-dependent, localized in space and extremely long-lived called
oscillons. It is investigated how the action of changing the minimum value
of the field configuration representing the oscillon affects its behavior.
We find that one of the consequences of this procedure, is the appearance of
a pair of oscillon-like structures presenting different amplitudes and
frequencies of oscillation. We also compare our analytical results to
numerical ones, showing excellent agreement.
\end{abstract}

\date{January 31, 2016}
\maketitle

\section{1. Introduction}

The presence of topologically stable configurations is an important feature
of a large number of interesting nonlinear models. Among other types of
nonlinear field configurations, there is a specially important class of
time-dependent stable solutions, the breathers appearing in the Sine-Gordon
like models. Another time-dependent field configuration whose stability is
granted for by charge conservation are the $Q$-balls \ as baptized by
Coleman \cite{S.Coleman1} or nontopological solitons \cite{Lee}. However,
considering the fact that many physical systems interestingly may present a
metastable behavior, a further class of nonlinear systems may present a very
long-living configuration usually known as oscillon. \ This class of
solutions was discovered in the seventies of the last century by Bogolyubsky
and Makhankov \cite{I.L.Bogolyubsky}, and rediscovered posteriorly by
Gleiser \cite{Gleiser1}. Those solutions, appeared in the study of the
dynamics of first-order phase transitions and bubble nucleation. Since then,
an increasing number of works have been dedicated to the study of these
objects \cite{Gleiser1}-\cite{PRL105(2010)}.

Oscillons are quite general configurations and are found in the
Abelian-Higgs $U(1)$ models \cite{Gleiser8}, in the standard model $%
SU(2)\times U(1)$ \cite{Graham2}, in inflationary cosmological models \cite%
{Gleiserintj}, in axion models \cite{Kolb}, in expanding universe scenarios 
\cite{Graham1, Mustafa} and in systems involving phase transitions \cite%
{Gleiser2}. In a recent work by Gleiser \textit{et. al.} \cite{Gleiser11}
the problem of the hybrid inflation characterized by two real scalar fields
interacting quadratically was analyzed. In that reference the authors have
shown that a new class of oscillons arise both in excited and ground states. 
\textcolor{black}{Here, it is important to
remark that an earlier mention of composite oscillons both in excited and
ground states was given in \cite{evangelos}, where have been obtained
oscillons in the $SU(2)$ Gauged Higgs Model (GHM).}

The usual oscillon aspect is typically that of a bell shape which oscillates
sinusoidally. Recently, Amin and Shirokoff \cite{Mustafa} have shown that
depending of the intensity of the coupling constant of the self-interacting
scalar field, it is possible to observe oscillons with a kind of plateau at
its top. In fact, they have shown that these new oscillons are more robust
against collapse instabilities in three spatial dimensions. In a recent work
the impact of the Lorentz and CPT breaking symmetries was discussed in the
context of the so-called flat-top oscillons \cite{AHEPRafael}.

At this point it is interesting to remark that Segur and Kruskal \cite%
{Segur1} have shown that the asymptotic expansion do not represent in
general an exact solution for the scalar field, in other words, it simply
represents an asymptotic expansion of first order in $\epsilon $, and it is
not valid at all orders of the expansion. They have also shown that in one
spatial dimension they radiate \cite{Segur1}. In a recent work, the
computation of the emitted radiation of the oscillons was extended for the
case of two and three spatial dimensions \cite{Gyula2}. Another important
result was put forward by Hertzberg \cite{Hertzberg}. In that work he was
able to compute the decaying rate of quantized oscillons, showing that the
quantum rate decay is very distinct from the classical one.

Thus, in this work we introduce novel configurations in one-dimensional
scalar field theories, which are time-dependent, localized in space and
extremely long-lived like the oscillons. This is done through the
investigation of how displacements of its position in the field potential
affects the features of the oscillons.

This paper is organized as follows. In section 2 we present the symmetrical
model which will be analyzed and we show the essential idea of the work. In
section 3 we find the respective oscillons-like configurations which we will
baptize as "Phantom oscillons". In section 4 we address to the problem of
the emitted radiation of the phantom oscillons. Discussion of some physical
features of the solutions are presented in section 5.

\section{2. The Basics}

In this work we study a real scalar field theory in $1+1$ space-time
dimensions described by the following action%
\begin{equation}
\mathcal{S}=\int dtdx\left[ \frac{1}{2}(\partial _{t}\phi )^{2}-\frac{1}{2}%
(\partial _{x}\phi )^{2}-V(\phi )\right] ,  \label{1}
\end{equation}

\noindent where $\partial _{t}=\partial /\partial t$, $\partial
_{x}=\partial /\partial x$, and $V(\phi )$ is the field dependent potential.
Thus, from the variation of the above action, the corresponding classical
field equation of motion can be written as%
\begin{equation}
\frac{\partial ^{2}\phi }{\partial t^{2}}-\frac{\partial ^{2}\phi }{\partial
x^{2}}=-\frac{\partial V(\phi )}{\partial \phi }.  \label{2}
\end{equation}

In order to introduce the idea we are pursuing, we will analyze the case of
the symmetric $\phi ^{6}$ model, which has been used in several contexts
devoted to the study of oscillons. Then, in this work, the model that we
will choose is represented by the following field potential%
\begin{equation}
V(\phi )=\frac{\lambda }{2}\phi ^{2}(a-b\phi ^{2})^{2}.  \label{3}
\end{equation}

\noindent where $\lambda $, $a$ and $b$ are real positive valued parameters.

The profile of this potential is illustrated in Fig. 1. This figure shows
that the potential presents three degenerate minima, localized in $\phi
_{v}^{(0)}=0$ and $\phi _{v}^{(\pm )}=\pm \sqrt{a/b}$.

\begin{figure}[h]
\centering
\includegraphics[width=8.0cm]{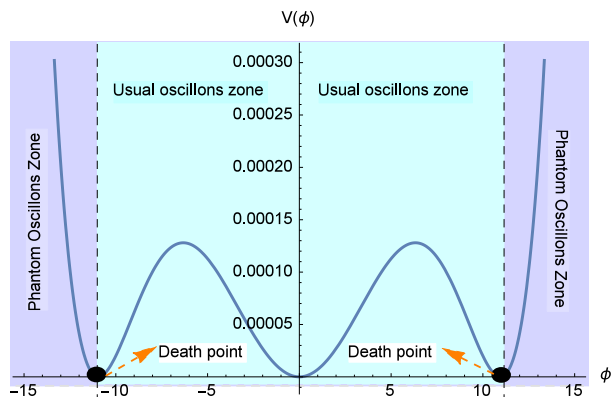}
\caption{{}Profile of the potential of the model on analysis with $a=1.2$, $%
b=1$ and $\protect\lambda =1$.}
\label{teste:oscillons}
\end{figure}
This kind of potential, presents kink-like configurations interpolating
adjacent vacua. Here, however, we are looking for time-dependent field
configurations which are localized in the space.

Therefore, since our primordial interest is to find localized and periodic
solutions, it is useful, as usual in the study of oscillons, to introduce
the following scale transformation in $x$ and $t$%
\begin{equation}
y=\epsilon x,\text{ }\tau =\sqrt{1-\epsilon ^{2}}t\text{.}  \label{4}
\end{equation}

\noindent where $0<\epsilon <<1$. Under these transformations, the field
equation (\ref{2}) becomes%
\begin{equation}
(1-\epsilon ^{2})\frac{\partial ^{2}\phi }{\partial \tau ^{2}}-\epsilon ^{2}%
\frac{\partial ^{2}\phi }{\partial y^{2}}+\lambda (a^{2}\phi -4ab\phi
^{3}+3b^{2}\phi ^{5})=0.  \label{5}
\end{equation}

From the above equation, it is possible to find the usual oscillons which
are localized in the central vacuum $\phi _{v}^{(0)}=0$ of the model
described by the potential (\ref{3}). In this case, the classical scalar
field $\phi $ is spatially localized and periodic in time. The usual
procedure to obtain oscillons configurations in $1+1$ dimensions consists in
applying a small amplitude expansion of the scalar field $\phi $ in powers
of $\epsilon $ in the following form%
\begin{equation}
\phi (y,\tau )=\sum_{n=1}^{\infty }\epsilon ^{n}\phi _{n}(y,\tau )\text{.}
\label{5.1}
\end{equation}

However, in this work, we are interested in analyze how some simple
displacements in the above expansion may affect the configuration and
stability of the oscillons. Then, we are searching for a small amplitude
solution where we expand the scalar field $\phi $ in powers of $\epsilon $ as%
\begin{equation}
\phi (y,\tau )=k+\sum_{n=1}^{\infty }\epsilon ^{n}\phi _{n}(y,\tau )\text{.}
\label{6}
\end{equation}

Note that the above expansion differs from the usual treatment of oscillons,
since we have an additive term, which corresponds to a translation in the
field. Furthermore, we can recover the usual expansion setting $k=0$. In
fact, considering that the terms in the expansion oscillates in such a way
that the mean value of the field configuration is $k$ (in fact, as we are
going to see below, it will appear an effective $k_{eff}$), one can see that
this new oscillon can be at regions of the potential which are far from its
vacua. As a consequence of this, we expect that these new field
configurations should be more unstable than the usual oscillons located at
the vacua of the potential. However, as we are going to see below, they are
still considerably long-living configurations. Furthermore, we discovered
that, after a given "death point", these oscillon type configurations are
separated from their "phantoms". This is going to be done in the next
section.

It is important to remark that our approach is general and can be applied to
different nonlinear field theories in order to investigate oscillons
configurations. An special case is that one given by choosing $b=0$ in the
model (\ref{3}). In this case the model becomes linear and the solution
involves both the spatial and temporal part. The temporal part is
oscillatory, but the spatial one is not localized. Nevertheless the method
still can be applied to find the solution.

\section{3. Phantom oscillons}

In this section, we will \ derive the profile of the proposed oscillon type
configurations using the expansion given by (\ref{6}). Let us substitute
this expansion of the scalar field into the field equation (\ref{5}). Thus,
it is not difficult to conclude that one gets%
\begin{eqnarray}
&&\left. \lambda k(a^{2}-4abk^{2}+3b^{2}k^{4})+\epsilon \left( \frac{%
\partial ^{2}\phi _{1}}{\partial \tau ^{2}}+\Gamma _{0}^{2}\phi _{1}\right)
\right.  \notag \\
&&\left. +\epsilon ^{2}\left( \frac{\partial ^{2}\phi _{2}}{\partial \tau
^{2}}+\Gamma _{0}^{2}\phi _{2}+\Gamma _{1}^{2}\phi _{1}^{2}\right) +\epsilon
^{3}\left( \frac{\partial ^{2}\phi _{3}}{\partial \tau ^{2}}-\frac{\partial
^{2}\phi _{1}}{\partial y^{2}}-\frac{\partial ^{2}\phi _{1}}{\partial \tau
^{2}}\right. \right.  \notag \\
&&\left. \left. +\Gamma _{0}^{2}\phi _{3}+\Gamma _{2}^{2}\phi
_{1}^{3}+\Gamma _{3}^{2}\phi _{1}\phi _{2}\right) +...=0.\right.  \label{6.1}
\end{eqnarray}

\noindent where we define%
\begin{eqnarray}
\Gamma _{0}^{2} &=&\Gamma _{0}^{2}(\lambda ,k,a,b)\equiv \lambda
(a^{2}-12abk^{2}+15b^{2}k^{4}), \\
\Gamma _{1}^{2} &=&\Gamma _{1}^{2}(\lambda ,k,a,b)\equiv 6bk\lambda
(5bk^{2}-2a), \\
\Gamma _{2}^{2} &=&\Gamma _{2}^{2}(\lambda ,k,a,b)\equiv 2b\lambda
(15bk^{2}-2a), \\
\Gamma _{3}^{2} &=&\Gamma _{3}^{2}(\lambda ,k,a,b)=2\Gamma _{1}^{2}.
\end{eqnarray}

We note that the procedure of performing a small amplitude expansion shows
that the scalar field solution $\phi $ can be obtained from a set of scalar
fields which satisfy coupled nonlinear differential equations. This set of
differential equations are found by taking the terms in all orders of $%
\epsilon $ in the above equation. However, using the expansion (\ref{6})
which has an additive term, one gets a constant term in order $\mathcal{O}%
(1) $ as output in (\ref{6.1}). Aiming to assure that the field
configuration is still oscillating in time, we must now suppose that the
first term is such that 
\begin{equation}
\lambda k(a^{2}-4abk^{2}+3b^{2}k^{4})=\eta \epsilon ^{3},  \label{cor}
\end{equation}%
where $\eta $ is a real arbitrary constant. This condition means that in
order to find the values of $k$ we need to establish the values of $\epsilon 
$ and $\eta $. In other words, $k=k(\epsilon ,\eta )$. Note that the above
supposition can be arranged through higher powers of $\epsilon $. However,
in order to obtain a correct relation between both sides of (\ref{cor}), we
impose that the constant $\eta $ is a number with a magnitude of order $%
10^{0}$. Thus, only for convenience, we put both sides of relation (\ref{cor}%
) in the same order of magnitude choosing the coupling constant very small
(for $\eta \sim 1$, $\lambda \sim \epsilon ^{3}$). In fact, the imposition (%
\ref{cor}) guarantees that no undesirable mixing between the orders of the
nonlinear differential equations appear. The motivation of our choice,
arises from the simple fact that we are including a residual effect in order 
$\mathcal{O}(\epsilon ^{3})$, which can be considered as an important
contribution in the oscillon configuration, otherwise we have not a relevant
effect in the configuration. Thus, it becomes immediately clear that up to $%
\epsilon ^{3}$ the above supposition leads to 
\begin{eqnarray}
&&\left. \frac{\partial ^{2}\phi _{1}}{\partial \tau ^{2}}+\Gamma
_{0}^{2}\phi _{1}=0,\frac{\partial ^{2}\phi _{2}}{\partial \tau ^{2}}+\Gamma
_{0}^{2}\phi _{2}=-\Gamma _{1}^{2}\phi _{1}^{2},\right.  \label{9} \\
&&\left. \frac{\partial ^{2}\phi _{3}}{\partial \tau ^{2}}-\frac{\partial
^{2}\phi _{1}}{\partial y^{2}}-\frac{\partial ^{2}\phi _{1}}{\partial \tau
^{2}}+\Gamma _{0}^{2}\phi _{3}+\Gamma _{2}^{2}\phi _{1}^{3}+\Gamma
_{3}^{2}\phi _{1}\phi _{2}=-\eta .\right.  \label{11}
\end{eqnarray}

Here, it is necessary to impose that $\Gamma _{0}$ has a real value,
otherwise the solution of $\phi _{1}(x,t)$ will not be oscillatory in time
and we will not obtain oscillon configurations. Therefore, from this
condition, we must impose that $a^{2}-12abk^{2}+15b^{2}k^{4}>0$. Now, under
this restriction, we have only a few acceptable ranges of validity {}{}for $%
k $, but we will see below that they will be enough to reproduce all the
values of the effective translation in the field. Furthermore, we will see
later that this restriction is not unique to get the possible values of the
translation constant.

Let us now look for the solution of the equations (\ref{9}) and (\ref{11}).
First, it is not difficult to conclude that, for $\Gamma _{0}$ real, the
solution of the equation (\ref{9}) can be given by 
\begin{equation}
\phi _{1}(y,\tau )=\varphi (y)\cos (\Gamma _{0}\tau ).  \label{12}
\end{equation}

Looking at the right equation of (\ref{9}), we see that the solution of $%
\phi _{2}(y,\tau )$ can be found by using the above solution. Thus,
substituting (\ref{12}) into the equation of $\phi _{2}$, one obtains%
\begin{equation}
\phi _{2}(y,\tau )=-\frac{\Gamma _{1}^{2}\varphi ^{2}(y)[3-\cos (2\Gamma
_{0}\tau )]}{6\Gamma _{0}^{2}}.  \label{13}
\end{equation}

Similarly, from the solutions (\ref{12}) and (\ref{13}) we can obtain $\phi
_{3}(y,\tau )$. Then, after straightforward calculations, one can verify
that the equation (\ref{11}) takes the form%
\begin{eqnarray}
&&\left. \frac{\partial ^{2}\phi _{3}}{\partial \tau ^{2}}+\Gamma
_{0}^{2}\phi _{3}=-\eta +\left[ \frac{d^{2}\varphi }{dy^{2}}-\Gamma
_{0}^{2}\varphi \right. \right.  \notag \\
&&\left. \left. +\left( \frac{5\Gamma _{1}^{2}}{6\Gamma _{0}^{2}}-\frac{%
3\Gamma _{0}^{2}}{4}\right) \varphi ^{3}\right] \cos (\Gamma _{0}\tau
)\right.  \label{14.2.2} \\
&&-\left( \frac{\Gamma _{2}^{2}}{4}+\frac{\Gamma _{1}^{2}}{6\Gamma _{0}^{2}}%
\right) \varphi ^{3}\cos (3\Gamma _{0}\tau ).  \notag
\end{eqnarray}

Our primordial goal is to get configurations which are periodical in time.
Then, if we solve the above partial differential equation in the presented
form, we will have a term linear in $\tau $. As a consequence, the solution
for $\phi _{3}$ is neither periodical, nor localized. This result comes from
the contribution of the function $\cos (\Gamma _{0}\tau )$ in the right-hand
side of the partial differential equation (\ref{14.2.2}). However, we can
construct solutions for $\phi _{3}$ which are periodical in time if we
impose that%
\begin{equation}
\frac{d^{2}\varphi }{dy^{2}}=\Gamma _{0}^{2}\varphi -\Omega _{0}^{2}\varphi
^{3},\text{ with }\Omega _{0}^{2}\equiv \frac{5\Gamma _{1}^{4}}{6\Gamma
_{0}^{2}}-\frac{3\Gamma _{2}^{2}}{4}.  \label{16}
\end{equation}

At this point it is necessary to impose that $\Omega _{0}^{2}>0$ in order to
find solutions with profile of hyperbolic secant. We recall from our studies
that $\Gamma _{0}$ is characterized by another condition, then we can now
combine the inequalities in order to encounter the valid region of the
values of $k$. Here, we shall not concern with such analytic detail. In
fact, such region is easily obtained in a numerical context, for instance 
\textcolor{black}{with $\lambda =10^{-5}$, $a=1.2$, $b=0.01$, and $\epsilon
=0.01$, we find $-3.3674<k<3.3674$, $k<-9.20112$ and $k>9.20112$.} Despite
this restriction, we will see later that these values {}{}will produce all
possible translations from the vacuum located at $\phi _{v}^{(0)}=0$. As a
consequence, we can choose the field configuration to be at any region of
the potential (\ref{3}).

Now, coming back to the equation (\ref{16}) and solving it, we obtain%
\begin{equation}
\varphi (y)=\frac{\Gamma _{0}\sqrt{2}\sech(\Gamma _{0}y)}{\Omega _{0}},
\label{17.3}
\end{equation}

Thus, using the condition that $\phi _{3}$ should be %
\textcolor{black}{localized}, the equation (\ref{14.2.2}) becomes%
\begin{equation}
\frac{\partial ^{2}\phi _{3}}{\partial \tau ^{2}}+\Gamma _{0}^{2}\phi
_{3}=-\eta -\left( \frac{\Gamma _{2}^{2}}{4}+\frac{\Gamma _{1}^{2}}{6\Gamma
_{0}^{2}}\right) \varphi ^{3}\cos (3\Gamma _{0}\tau ),
\end{equation}

\noindent which has the corresponding solution%
\begin{equation}
\phi _{3}(y,\tau )=-\frac{\eta }{\Gamma _{0}^{2}}+\frac{\omega
_{0}^{2}\varphi ^{3}\cos (3\Gamma _{0}\tau )}{8\Gamma _{0}^{2}},\text{ with }%
\omega _{0}^{2}\equiv \frac{\Gamma _{2}^{2}}{4}+\frac{\Gamma _{1}^{2}}{%
6\Gamma _{0}^{2}}.
\end{equation}

From the above results, as one can see, up to order $\mathcal{O}(\epsilon
^{3})$, the corresponding solution for the classical field is given by%
\begin{eqnarray}
&&\left. \phi (y,\tau )=k_{eff}+\epsilon \varphi (y)\cos (\Gamma _{0}\tau
)+\epsilon ^{2}\frac{\Gamma _{1}^{2}\varphi ^{2}(y)[\cos (2\Gamma _{0}\tau
)-3]}{6\Gamma _{0}^{2}}\right.  \notag \\
&&\left. +\epsilon ^{3}\frac{\omega _{0}^{2}\varphi ^{3}(y)\cos (3\Gamma
_{0}\tau )}{8\Gamma _{0}^{2}}+\sum_{n=4}^{\infty }\epsilon ^{n}\phi
_{n}(y,\tau ),\right.
\end{eqnarray}

\noindent with $k_{eff}\equiv
4bk^{3}(3bk^{2}-2a)/(a^{2}-12abk^{2}+15b^{2}k^{4})$, which is the
corresponding effective mean value of the field configuration as mentioned
in the second section. Since the parameter $\epsilon $ is taken as extremely
small, the profile of the solution is defined up to order $\mathcal{O}%
(\epsilon )$, once that subsequent orders will be even smaller. Thus, the
field \ configuration written in terms of the original variables, in a good
approximation, is given by 
\begin{equation}
\phi _{osc}(x,t)\approx k_{eff}+\epsilon \left[ \frac{\sqrt{2}\Gamma _{0}%
\sech(\epsilon \Gamma _{0}x)}{\Omega _{0}}\right] \cos (\Gamma _{eff}t)+%
\mathcal{O}(\epsilon ^{2}),
\end{equation}

\noindent with $\Gamma _{eff}=\Gamma _{0}\sqrt{1-\epsilon ^{2}}$.

We can note that the fundamental difference of our solution, compared to the
usual oscillon, is given by of the presence of $k_{eff}$. Here we have the
advantage of moving the configuration for any region within the potential,
this important mechanism opens a window to show that it is possible to find
oscillons living in any region of the potential, but a natural question
arises about the stability of this configuration, once that the usual
oscillons are highly stable configurations when they are oscillating around
the vacuum of the potential. In the next section we will address this
question about the stability of the our solution and we will see that these
oscillons are more unstable than the ones localized in the vacuum.

Another important and interesting result that arises from of our solution is
the emergence of a new "phantom oscillon" after a certain threshold value of 
$k_{eff}$. Above that value, what we called a "death point" we will have
always the presence of two oscillons, in a kind of "phantom zone" (see Fig.
1).

In Fig. 2, we sketch of the typical profile of the oscillon and its phantom.
In that figure, we see that the field configurations are oscillating around
of the corresponding effective mean value of the field configuration.
Furthermore, we can note that both the real oscillon as its phantom have
different amplitudes in their structures located at the origin. Moreover,
the tail of the phantom reaches a value of the effective field in a spatial
region farther than the real oscillon.

\begin{figure}[h]
\centering
\includegraphics[width=9.8cm]{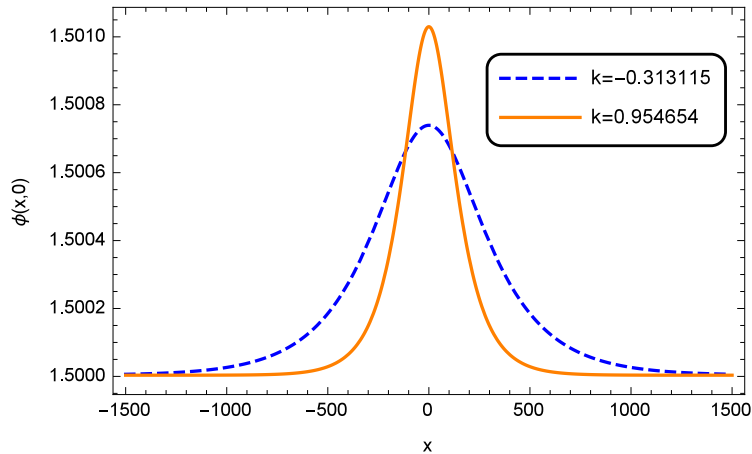}
\caption{{}Oscillon (thin line) and its phantom (dashed line) for $t=0$, $%
\protect\epsilon =0.01$, $a=1.2$, $\protect\lambda=10^{-5}$ and $b=0.01$.}
\label{Fig2:oscillons}
\end{figure}

We also emphasize that in the phantom zone, the effective frequency $\Gamma
_{eff}$ of the oscillations in time for the oscillons decrease with $k_{eff}$%
. In Fig. 3 we plot $\Gamma _{eff}$ as a function of $k_{eff}$. There one
can see that the oscillon frequency is decreasing with $k_{eff}$ and that
the phantom oscillon has a higher frequency than that of the corresponding
oscillon for a given value of $k_{eff}$.

\begin{figure}[h]
\centering
\includegraphics[width=9.2cm]{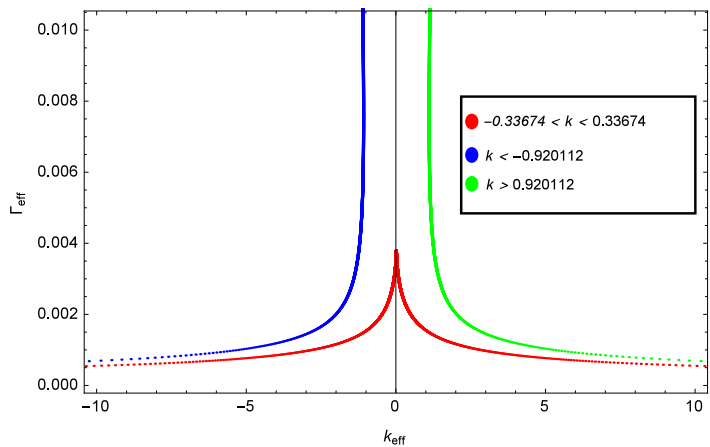}
\caption{{}Effective frequency of the oscillon (bottom line) and its phantom
(top line) for $t=0$, $\protect\lambda =10^{-5}$, $\protect\epsilon =0.01$, $%
a=1.2$ and $b=0.01$.}
\label{Fig6:oscillons}
\end{figure}

\begin{figure}[h]
\centering
\includegraphics[width=9.2cm]{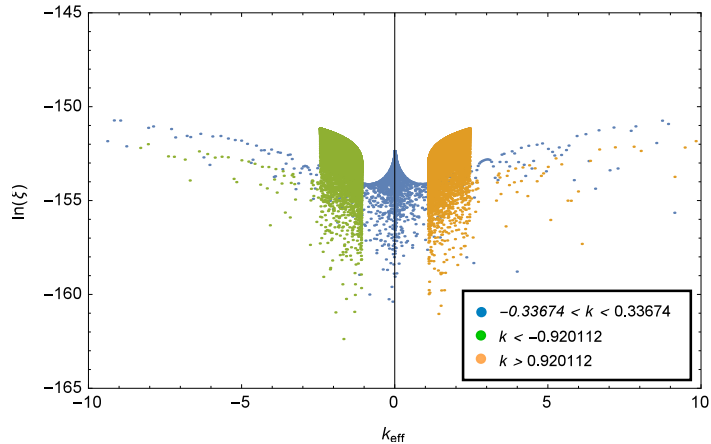}
\caption{{}Radiation power for $t=0$, $x=250$, $\protect\lambda =10^{-5}$, $%
\protect\epsilon =0.01$, $a=1.2$ and $b=0.01$.}
\label{Fig7:oscillons}
\end{figure}

\section{4. Radiation}

An important characteristic of the oscillons is its radiation emission,
responsible for its unavoidable decaying. In a seminal work by Segur and
Kruskal \cite{Segur1}, it was shown that oscillons in one spatial dimension
decay emitting radiation. Recently, the computation of the emitted radiation
in two and three spatial dimensions was did in \cite{Gyula2}. On the other
hand, in a recent paper by Hertzberg \cite{Hertzberg}, it was found that the
quantum radiation is very distinct from the classic one. It is important to
remark that the author has shown that the amplitude of the classical
radiation emitted can be found using the amplitude of the Fourier transform
of the spatial structure of the oscillon.

Thus, in this section, we compute the outgoing radiation of these oscillon
type configurations. Here, we will use a method in $1+1$ dimensional
Minkowski space-time that allows to compute the classical radiation, and
which is similar to the one presented in \cite{Hertzberg}. This approach
supposes that we can write the solution of the classical equation of motion
in the following form%
\begin{equation}
\phi _{sol}(x,t)=\phi _{osc}(x,t)+\xi (x,t),  \label{r1}
\end{equation}

\noindent where $\phi _{osc}(x,t)$ is the oscillon solution and $\xi (x,t)$
represents a small perturbation. Let us substitute this decomposition of the
scalar field into the equation of motion (\ref{2}). This leads to%
\begin{equation}
\frac{\partial ^{2}\xi }{\partial t^{2}}-\frac{\partial ^{2}\xi }{\partial
x^{2}}+\Lambda (x,t)\xi =-J(x,t),  \label{r2}
\end{equation}

\noindent where $\Lambda (x,t)=\lambda (a^{2}-12ab\phi
_{osc}^{2}+15b^{2}\phi _{osc}^{4})$ and the function $J(x,t)$ acts as an
external source. In this case it is written as 
\begin{equation}
J(x,t)=\frac{\partial ^{2}\phi _{osc}}{\partial t^{2}}-\frac{\partial
^{2}\phi _{osc}}{\partial x^{2}}+\lambda (a^{2}\phi _{osc}-4ab\phi
_{osc}^{3}+3b^{2}\phi _{osc}^{5}).  \label{r2.1}
\end{equation}

Since $\xi $ represents a small correction, we naturally assume that the
dependence of the nonlinear terms in higher powers by $\xi $ can be
neglected. On the other hand, we will look for solutions of $\xi (x,t)$
where the amplitudes of the tails of the oscillons are very smaller than
those of the core. In other words, we are looking for solutions at large
distances. This says that for $x>>1$ the field configuration is given by $%
\phi _{osc}\simeq k_{eff}+\epsilon \varphi (x)\cos (\Gamma _{eff}t)$. Thus,
the equation (\ref{r2}) can be rewritten as%
\begin{equation}
\frac{\partial ^{2}\tilde{\xi}}{\partial t^{2}}-\frac{\partial ^{2}\tilde{\xi%
}}{\partial x^{2}}+\tilde{\Lambda}(x,t)\tilde{\xi}=-\tilde{J}(x,t),
\label{r2.2}
\end{equation}%
where 
\begin{eqnarray}
&&\left. \tilde{\xi}(x,t)=\xi (x,t)+\mu _{eff},\tilde{\Lambda}(x,t)\simeq
\gamma _{0}\sech(\epsilon \Gamma _{0}x)\cos (\Gamma _{eff}t),\right.
\label{r2.3} \\
&&\left. \tilde{J}(x,t)\simeq -\frac{\sqrt{2}\Gamma _{0}^{3}}{\Omega _{0}}%
\sech(\epsilon \Gamma _{0}x)\cos (\Gamma _{eff}t)\right.  \notag \\
&&\left. +\left( \epsilon \tilde{A}+\epsilon ^{2}\frac{2\sqrt{2}\Gamma
_{0}^{2}}{\Omega _{0}}\right) \sech(\epsilon \Gamma _{0}x)\cos (\Gamma
_{eff}t),\right. \\
&&\left. \mu _{eff}\equiv \frac{a^{2}k_{eff}-4abk_{eff}^{3}+3b^{2}k_{eff}^{5}%
}{a^{2}-12abk_{eff}^{2}+15b^{2}k_{eff}^{4}},\right. \\
&&\left. \gamma _{0}\equiv -12A\epsilon \lambda
bk_{eff}(2a-5bk_{eff}^{2}),\right. \\
&&\left. \tilde{A}\equiv \frac{\sqrt{2}\Gamma _{0}a\lambda (a-12bk_{eff}^{2})%
}{\Omega _{0}}.\right.
\end{eqnarray}

At this point we can obtain from the equation (\ref{r2.2}), that%
\begin{equation}
\tilde{\xi}(x,t)=-\frac{1}{(2\pi )^{2}}\lim_{p\rightarrow 0^{+}}\int d\bar{%
\omega}\int d\bar{k}\frac{J(\bar{\omega},\bar{k})e^{i(\bar{k}x-\bar{\omega}%
t)}}{\bar{k}^{2}-\bar{\omega}^{2}\pm ip},  \label{r6.1}
\end{equation}

\noindent with%
\begin{equation}
J(\bar{\omega},\bar{k})=\int d\bar{x}\int d\bar{t}J(\bar{x},\bar{t})e^{-i(%
\bar{k}x-\bar{\omega}t)}.  \label{r6.2}
\end{equation}

It is important to remark that in order to solve analytically the equation (%
\ref{r2.2}), we will impose that $b$ is of the order of $\mathcal{O}%
(\epsilon )$, as a consequence $\gamma _{0}\sim \epsilon ^{2}$. Therefore,
at far distances from the core of the oscillon, we have $\tilde{\Lambda}%
(x,t)\sim 0$. Thus, after straightforward computations, one can conclude that%
\begin{equation}
\tilde{\xi}(x,t)\simeq \frac{\sqrt{2}J_{0}\sqrt{\pi }}{4\epsilon \Gamma
_{0}\Gamma _{eff}}\sech\left( \frac{\Gamma _{eff}\pi }{2\epsilon \Gamma _{0}}%
\right) \sin (\Gamma _{eff}x)\cos (\Gamma _{eff}t),  \label{r8}
\end{equation}

\noindent where $J_{0}=-\sqrt{2}\Gamma _{0}^{3}/\Omega _{0}+\epsilon \tilde{A%
}+2\sqrt{2}\epsilon ^{2}\Gamma _{0}^{3}/\Omega _{0}$. 
\textcolor{black}{Using the Eq. (\ref{r8}), we can see in Fig. 4 the profile
of the radiation power emitted,} where the reader can observe that the
stability of the dislocated oscillon diminishes for increasing values of $%
k_{eff}$, and that the phantom oscillon located at the second vacuum is as
stable as the oscillon located at the central vacuum.

\section{5. Numerical results}

In this section, in order to check our analytical results, we will compute
the numerical solutions for the oscillon profile. In this way, to analyse
the oscillon configuration numerically we use the initial configuration in
the form%
\begin{equation}
\phi _{num}(x,0)=k_{eff}+\epsilon \frac{\Gamma _{0}\sqrt{2}\sech(\Gamma
_{0}\epsilon x)}{\Omega _{0}}.
\end{equation}

In general oscillons are do not an exact solution for the scalar field.
Thus, it is convenient to begin with the above initial configuration for
evolution of the numerical \textcolor{black}{solution}. Another important
condition is given by $\partial \phi (x,0)/\partial t=0$.

Now, for evolution of the numerical solution of the field equation (\ref{5})
we will use $\epsilon =0.01$, $k=-0.1$, $a=1.2$, $b=1$ and $\lambda =1$. As
a result of this choise, ones have $k_{eff}\simeq 0.07$. To illustrate the
field configurations which corresponds to the oscillons we graph the field
at $x=0$ and a density plot of the field $\phi (x,t)$. We can see this
numerical field solution in Fig. 5. That figure shows a comparison of
analytical solution and numerical ones. We can observe that the analytical
value of the field at the centre of the oscillon shows a small disparity in
the horizontal position when the time increases showing a profile of phase
difference. However, looking in a large range of the time we can find that
the disparity is of order $\sim 6\%$ showing, as we would expect, that the
theoretical solutions are in good agreement with the numerical ones.
Furthermore, in the vertical position the analytical amplitude of the
oscillons are correctly predicted by the numerical results.

\begin{figure}[h]
\centering\includegraphics[width=8.8cm]{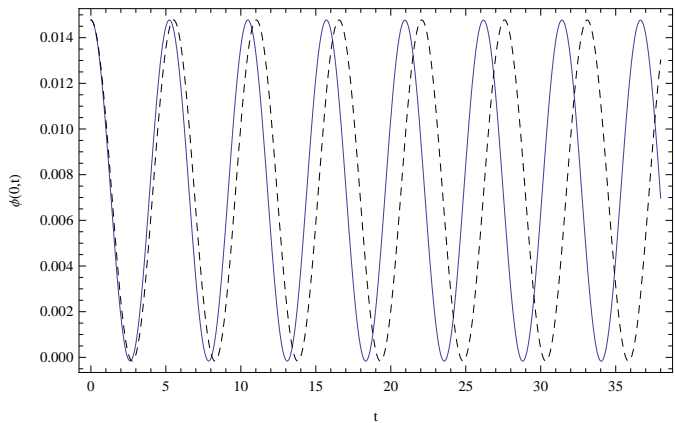} %
\caption{Analytical and numerical results. The figure is a comparison of the
field at $x=0$. Dashed line is analytical and \ the solid curve is
numerical. }
\label{numerical}
\end{figure}

\section{6. Conclusions}

In this work we have presented novel oscillon-like configurations which we
call phantom oscillons. We have found that displacements of the minimum
value of the field configuration representing the oscillon affects its
behavior. In this case, the procedure of introducing those displacements act
as a kind of source of new oscillons, in fact this leads us to think about
the possibility of the appearance of a higher number of additional oscillons
when one deals with a field potential containing a bigger number of
degenerate vacua. This possibility is under analysis and hopefully will be
reported in a future work. Moreover, it can be observed in the Figure 4 that
the original oscillon (created at $\phi _{vac}=0$) is more stable than the
ones located at ($\phi _{vac}=k_{eff}$) and the stability decreases when $%
k_{eff}$ increases. This happens until one reaches the position of one
another vacuum of the model. At this point a second oscillon-like
configuration shows up (the phantom oscillon), and it is remarkable that it
presents approximately the same degree of stability that of the original
oscillon. Furthermore, one can also note that the frequency of the phantom
oscillon is always higher than the corresponding oscillon (see Figure 3). An
interesting consequence of these configurations is that one can think about
the behavior of a gas of oscillons, where a statistical distribution of
oscillons would appear \cite{gleisernew}, each one having a different value
of $k_{eff}$ and, due to the relative stabilities, some of them would decay
or combine to produce more stable structures. This kind of scenario could be
of interest for some cosmological models \cite{mustafa2, mustafa3, mustafa4}.

\bigskip

\noindent \textbf{Acknowledgments}

The authors thanks CNPq and CAPES for partial financial support. R. A. C. C.
also thank Marcelo Gleiser for helpful discussions and for valuable remarks
about oscillons. R. A. C. C. and A. S. D. also would like to thank the
anonymous referee for the valuable comments.

\end{document}